# Mapping the path to Cryogenic Atom Probe Tomography Analysis of biomolecules

Eric V. Woods[1, *], Tim M. Schwarz[1], Mahander P. Singh[1], Shuo Zhang[1], Se-Ho Kim[1,2], Ayman A. El-Zoka[1,3], Lothar Gremer[4,5], Dieter Willbold[4,5], Ingrid McCarroll[1], and B. Gault[1,3, *]

1. Max-Planck-Institute for Sustainable Materials, 40237 Düsseldorf, Germany

2. now at Department of Materials Science and Engineering, Korea University, Seoul 02841, Republic of Korea

3. Department of Materials, Royal School of Mines, Imperial College London, London, UK.

4. Institute of Biological Information Processing, Structural Biochemistry (IBI-7), Forschungszentrum Jülich, 52425 Jülich, Germany.

5. Institut für Physikalische Biologie, Heinrich-Heine-Universität Düsseldorf, 40225 Düsseldorf, Germany.

*Corresponding Authors: e.woods@mpie.de, b.gault@mpie.de

# Abstract

The understanding of protein structure, folding, and interaction with other proteins remains one of the grand challenges of modern biology. Tremendous progress has been made thanks to X-ray- or electron-based techniques that have provided atomic configurations of proteins, and their solvation shell. These techniques though require a large number of similar molecules to provide an average view, and lack detailed compositional information that might play a major role in the biochemical activity of these macromolecules. Based on its intrinsic performance and recent impact in materials science, atom probe tomography (APT) has been touted as a potential novel tool to analyse biological materials, including proteins. However, analysis of biomolecules in their native, hydrated state by APT have not yet been routinely achieved, and the technique's true capabilities remain to be demonstrated. Here, we present and discuss systematic analyses of individual amino-acids in frozen aqueous solutions on two different nanoporous metal supports across a wide range of analysis conditions. Using a ratio of the molecular ions of water as a descriptor for the conditions of electrostatic field, we study the fragmentation and behavior of those amino acids. We discuss the importance sample support, specimen preparation route, acquisition conditions and data analysis, to pave the way towards establishing guidelines for cryo-APT analysis of biomolecules.

# 1 Introduction

Since proteins and biomolecules are fundamental to life, the analysis of their interactions and configuration is crucial for understanding biological functionality. Primary protein structure consists of an initially linear chain of twenty canonical amino acids, which share a common backbone in their free state. This backbone contains a carboxyl group and an amine group, which are subsequently chemically joined to form a so-called peptide bond during synthesis [1]. The variable portion is a side group (termed 'R'), which consists of a wide variety of different chemical configurations, chain length, and positive and negative charges, e.g. exposed hydroxyl and amine groups, which can serve as bonding sites for enzymes to later perform post-translational modification (PTM) binding [2,3]. As the protein emerges from the ribosome during synthesis, the side groups of the protein drive folding into certain common motifs, such as α-helices and β-sheets, which comprise the secondary structure [1]. Other cellular machinery, such as chaperone proteins, may assist the nascent protein to achieve its stable larger, three-dimensional ternary form [4]. Finally, the folded protein may be assembled into larger macromolecular complexes, so-called quaternary structures, including respiratory complexes, nuclear pores, etc [5].

As a part of this process, the protein may be complexed with metal ions, such as iron and magnesium, to catalyse functions such as oxygen transport, chemical transformation, and a whole panoply of other reactions essential to life that require co-factors to perform [6]. Characterizing the interaction of metal ions with proteins, for example, is an exceptionally challenging process in the dynamic, ever-changing environment of a cell. These fleeting interactions are critical to life processes, and known to generate reactive oxygen species (ROS), and both copper and iron are implicated in Alzheimer's disease and many other neurological disorders [7]. Studying the structure of native proteins has been revolutionized by the availability of cryogenic transmission electron microscopy (cryo-TEM) [8–10], but it has significant challenges in elemental identification for metal ion interactions, with cryo-scanning TEM (cryo-STEM) being well positioned to overcome this to a limited extent [11,12]. For studying proteins' chemistry in their native state, tandem mass-spectrometry (MS) is the other state-of-the-art technique [13]. The two techniques provide complementary information but on scales that are difficult to reconcile, failing to provide full 3D elemental identification and positional information at a sub-nanometre scale.

Atom probe tomography (APT) can achieve elemental and ionic mapping in three dimensions (3D) with sub-nanometre spatial resolution [14,15]. APT's potential to address many open, outstanding biological questions has been discussed for decades [16–18], but studies beyond proofs-of-principle are rare and focused primarily on hard biominerals [19–21].

During APT analysis, a sharp needle-shaped specimen is maintained at cryogenic temperatures and subjected to a high DC electrical field on which are superimposed either voltage pulses [22], laser pulses [23,24], or a combination of both [25,26]. The pulses drive the field evaporation of individual atomic or molecular ions, which are collected by a position-sensitive, time-resolved detector, recording simultaneously a 2D ion impact position and a time-of-flight converted into a mass-to-charge ratio [27,28]. This data is processed to produce a three-dimensional point cloud reconstituting the distribution of species in the original specimen [29].

APT specimens are typically less than 100 nm in diameter, and are usually prepared using a dual-beam scanning electron microscope / focused-ion beam (SEM/FIB) systems (Prosa & Larson, 2017). For wet or hydrated samples, preparation has historically been extremely challenging. Ideally, to avoid frost build-up, handling of frozen samples must be carried out in inert gas gloveboxes either before or after the liquid sample is frozen. Cryogenic ultra-high-vacuum (UHV) transfer shuttles facilitate sample transfer between instruments. Several groups have developed integrated cryogenic instrumental suites and workflows [30–32] or built coupled FIB/SEM-APT systems [33]. Specimen preparation has been reported from the edge of frozen droplet on nanoporous flat substrates [34], nanoporous metal needles [35] or from very large droplets placed on wire-shaped substrates [36,37]. Site-specific liftout of a 2-3 μm thick lamella at cryogenic temperature, subsequently mounted on APT multi-needle specimen carrier was also demonstrated [38,39], and further refined to enable more routine cryogenic specimen preparation [40], including lift-out of frozen liquid water without any support material [41].

In order to analyse organic molecules using APT, particularly proteins embedded in water, both the fragmentation of model systems, e.g. single amino acids, and field evaporation of the aqueous matrix should be well characterized and understood. Individual amino acids, prepared via drop-casting, e.g. placing a drop of liquid, an amino acid solution onto a carbon nanotube (CNT) mesh and dried, were previously studied in APT [42–44], but without the matrix material the applicability to understanding hydrated proteins and their constituent amino acids is likely limited, both because the electric field conditions are likely significantly different and the amino acids probably ionize differently in zwitterionic form. Significant work has gone into understanding the field emission behaviour of water in conjunction with theoretical work including [45–49]. In APT specifically, pure water has been analysed in free-standing specimens [34,36,50], graphene-encapsulated needles [51,52], and on or in nanoporous metal substrates [35,53–56].

There is a clear need to advance the state-of-the-art. Here, we present here the results of an effort to characterize single amino acids in a water matrix on nanoporous metals, as model systems to understand their fragmentation patterns. Data is presented for cysteine, lysine, and arginine, on two substrates, nanoporous gold (NPG) and nanoporous copper (nanoporous Cu) from a CuZn alloy, with two different local-electrode atom probes (LEAP) and across a wide range of analysis conditions. This comprehensive set of data allows us to outline guidelines for the analysis of amino acids and more complex biomacromolecules that will help establish best practice for this burgeoning field.

## 2 Results

### 2.1 Water matrix

Field evaporation of organic, and generally non-metallic, materials leads to the formation of molecular or cluster ions that can dissociate into smaller fragments. The fragmentation path depends on the electric field conditions. This requires identifying a reliable desciptor to estimate of the electrical field strength across different water-containing datasets, in order to facilitate systematic analyses to understand, and possibly predict the results of the fragmentation process. For metallic systems, or semiconductors, the ratio of the charge-states of an element is often used a proxy for local electrical field [62–65]. Here, since water is the typical matrix for biomolecules, defining a descriptor associated to the ionic signals from the analysis of water appears be the most reasonable choice.

Preliminary work demonstrated that water field evaporated in the form of protonated water cluster ions with a formula $(H_2O)_nH^+$ [34,36,49], and their relative abundance is an indicator of the electric field strength. Here, we considered 20 individual datasets, some containing data acquired across a range of laser pulse energies, and the three most abundant ions were for n=1, 2, 3. At higher field strengths, higher order water clusters above n=3 are below the level of background. In addition, in most datasets the difference in abundance of the n=1 and n=2 varied within a similar, limited range. These observations agree with the previously mentioned references.

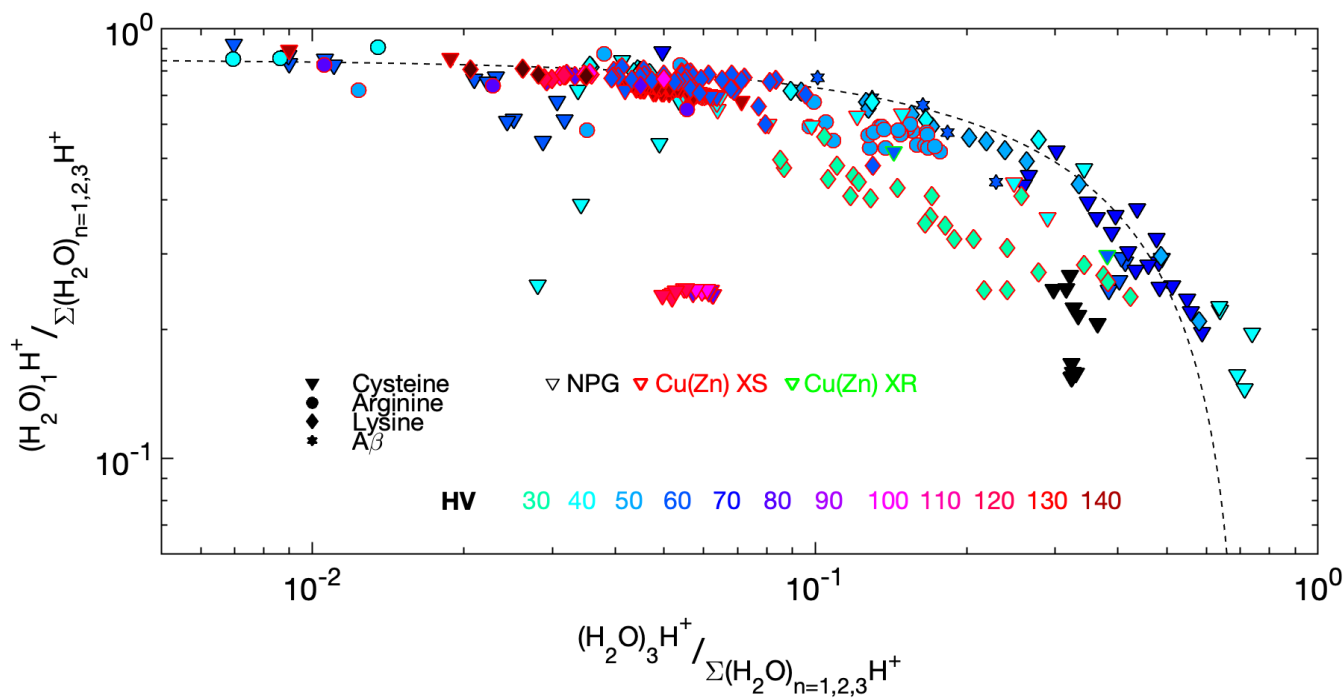

**Figure 1**. Plot of n=1, $(H_2O)H^+$, normalised by the sum of water clusters n=1, 2, 3, $(H_2O)_n H^+$, as a function of the n=3, $(H_2O)_3 H^+$, similarly normalised. The dashed line follows a linear relationship (y=0.85−1.2x) and is just a guide for the eye.

**Figure 1** plots logarithmically the relative abundance of the protonated water clusters $(H_2O)_1 H^+$ vs. $(H_2O)_3 H^+$, both normalised by $\sum(H_2O)_{n=1,2,3} H^+$, across the 20 datasets, split into subsets containing 1 million ions. The symbol represents the biomolecule in the solution, the colour represents the laser pulse energy, with black representing data acquired in high voltage pulsing mode. The symbol with a black border was obtained using NPG as a substrate, whereas the red and green border are data acquired with a nanoporous Cu substrate, respectively on a straight flight path (XS) or reflectron-fitted (XR) instrument. The dashed line is just a guide for the eye.

This graph is admittedly complex, but showcases general trends. Overall, the field conditions do not directly correlate with the instrument used or the laser pulse energy. This is expected because the thermal pulse arising from laser absorbed by the specimen depends on numerous factors [66–68], including, for example, the specimen's geometry, i.e. radius and shank angle [24]. These even vary over the course of a single analysis, including through a loss of cylindrical symmetry from the side first illuminated by the laser [69], along with material or structural inhomogeneities. Additionally, the UV laser absorption coefficient for water in an external electric field, along with any material in solution, must be taken into consideration and are mostly unknown. The data obtained in HV pulsing could have been expected to lead to the highest electric field conditions, yet recent reports indicate a more complex behaviour for the field evaporation of water under intense field conditions [36,50]. Some data (e.g. cysteine on Cu(Zn) in the XS, red triangles) are clearly off the main trend, which may be due to longer specimens[61] or, as will be discussed below, a large amount of metallic impurities.

Generally, though, the data aligns on a negative linear slope, which matches the expected behaviour from previous reports [46,47,50,70]. This near monotonous decrease is encouraging as it shows that the relative fraction of $(H_2O)_3H^+$ can inform on the electric field conditions. We can hence define a cluster ion ratio (CIR) as $\sum_{n=1}^{n=3}(H_2O)_n H^+/(H_2O)_3 H^+$, which takes values >1 and that can be considered a suitable descriptor for the electric field conditions, and which will be used hereafter. Practically, this means that datasets with a relatively higher CIR means that local electrical field is substantially higher than that of a dataset with a lower CIR. A low CIR will have more water clusters, $(H_2O)_nH^+$, where n is the cluster order, probably at least $n$ = 1…8. Previous work has illustrated that water clusters up to $n$ = 18 are possible [55,56,61]. A dataset with high CIR would be expected to suppress larger protonated water clusters, such that only few clusters over $n$= 3 are visible over background. With a lower CIR, the local electric field is lower and therefore larger clusters are energetically favourable [47].

## 2.2 Lysine

We first focus on lysine, with a chemical formula $C_6H_{13}N_2O_2$ and a mass of 146 Da. In **Figure 2(a)**, a mass spectrum obtained for an APT analysis of lysine in pure DI water is plotted, along with the chemical structure inset in the upper right. The substrate was NPG, but the specimen was finalised far from the metal substrate (3–5 µm). Salient features include (1) protonated water cluster ions, $(H_2O)_nH^+$, where $n$=1…9, labelled in blue, with their respective locations indicated the blue arrows; (2) organic fragments originating from lysine, i.e. that are not detected compared to the analysis of pure DI water, are labelled in red. Post-evaporation ionization can and does occur, and therefore multiple charged ions/molecules [71–74] can be detected, which further complicates the unambiguous identification of the signals. Secondly, larger molecules can dissociate during their flight to the detector and identifying their parent molecular ions is important for possible spatial reconstruction. This can be achieved using correlated ion evaporation histograms [75] generated by multiple hits on the detector created a single pulse. In such a correlated evaporation histogram, ion tracks which show horizontal and vertical lines corresponds to the field evaporation trigger by the laser or high voltage plus. Lines that are parallel to the bisector of the plot with positive slope demonstrate delayed evaporation, i.e. the evaporation of an ion after a certain amount of time after one ion was already field evaporation trigger by the laser/voltage pulse. implying they did not undergo post-evaporation ionization or field-induced dissociation or fragmentation of molecular or cluster ions during the flight towards the detector, which would appear as a curved line with negative slope [76]. A correlated evaporation histogram for the lysine in water analysis, shown in Figure **SI 1**, shows no indication of post-evaporation dissociation.

The fragments were identified based on likely breakdown pathways from literature including different electrospray mass spectroscopy techniques [77–80], along with APT analyses of single amino acid and dipeptide data in dried form [42–44,81]; single protein APT data [52]; and APT data of organic-rich portions of collagen in bone, which consist of proline or hydroxyproline, glycine, and some other amino acids [21,82–84] and the assumption to have only single-charged molecules. However, an unambiguous identification of the signals is difficult due to an overlap of certain protonated water cluster ions c with organic fragments, for instance, the protonated water cluster with $n$ = 8, $(H_2O)_8H^+$, has mass-to-charge 145 Da, precludes measurement of intact or protonated lysine that may have appeared. Additionally, certain masses are simply hard to unambiguously assign: drop-cast and dried lysine had a characteristic APT fragment signal, assigned as $CONH^+$ at 43 Da, with additional smaller peaks at 42 and 44 Da (likely de-protonated and protonated forms respectively) [42,44]. In the current work with aqueous solutions, the 44 Da peak is assigned as $CO_2^+$, but there can be contribution from $CONH_2^+$ and possibly $C_3H_8$, although those ions would be indistinguishable based on their mass-to-charge alone, presuming singly-charged ions & molecules. Across the references discussed previously, primarily the mass-to-charge 44 Da was assigned to $CO_2^+$ or $CONH_2^+$, where $C_3H_8^+$ was found in dried organics [43,59].

As shown in **Figure 2(b),** incomplete dealloying or cleaning of the NPG substrate can result in both Au and Ag dissolving into the solution. The Au is labelled in yellow and the Ag in grey. The correlated evaporation histogram in Figure **SI 2** shows no indication of post-evaporation dissociation. Water cluster ions with n=1–5 are observed, but any higher mass ones would be occluded by the metal peaks. Enlarged views of the mass spectral ranges where specific Ag and Au metal-water or metal-organic complexes form in the mass spectrum are displayed for Ag in **SI 3** and for Au in **SI 4** respectively, as previously reported [34]. These demonstrate that both Ag and Au form molecular and/or cluster ions complexed with $CO^+$ and/or $CO_2^+$ ions in aqueous matrices. While noting that the –CO group could also be -$CNH_3$, these ions were previously unreported in aqueous matrices, but indicates possible preferential bonding of the lysine fragments to the metallic ions in the solution. This behaviour could modify a particular amino acid's field evaporation behaviour, possibly complicating future attempts at quantification using APT. Practically, it means that when analysing ion locations (and visualizing, for example, protein backbones) on nanoporous metal substrates, mass peaks for specific metal-$CO^+$ and metal-$CO_2^+$ ions must be included in the analysis.

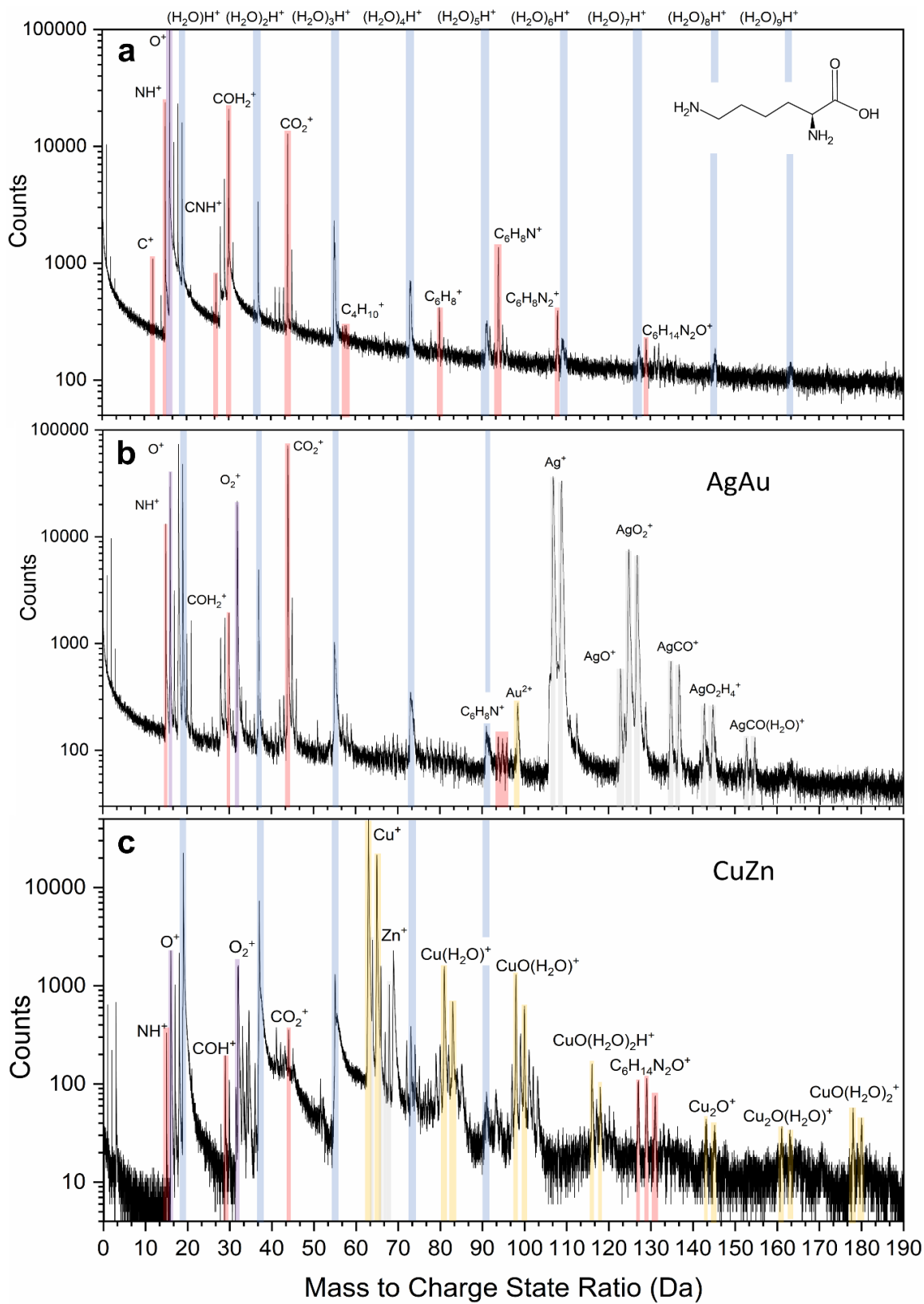

**Figure 2.** Mass spectra for lysine (a) in Type 1 ultrapure DI water on NPG, with no apparent metal (12.4 million ions, CIR: 0.26) (b) in water, Ag and Au ions from NPG substrate (6 million ions, CIR: 10.2), substrate name inset on right (c) in water, including Cu and Zn ions from CuZn metal substrate (3 million ions, CIR: 9.76), substrate name inset on right.

**Figure 2(c)** illustrates the same phenomena for Cu and Zn, although the Zn is almost completely removed from the substrate [56]; Cu peaks are labelled in yellow, Zn is labelled as grey. The Cu peaks dominate the mass spectra, shown in more detail in **SI 5**, and make it extremely difficult to see any organic fragments from lysine. There are mass peaks associated with hydrated Cu oxides which are labelled. In this mass spectrum, none of their individual peak heights exceeds 5.8% of the $^{63}$Cu$^+$ peak height. However, in other mass spectrums as discussed later, the height of those Cu oxide and water complexes can significantly exceed the $^{63}$Cu$^+$ peak height. The signal from Cu and their oxide species make it difficult to see any higher ordered protonated water clusters above *n* = 3. Different from the other correlation histograms, **SI 6** clearly shows ion tracks with negative slope, which correspond to the post-evaporation dissociations as discussed in the accompanying **SI**.

## 2.3 Arginine

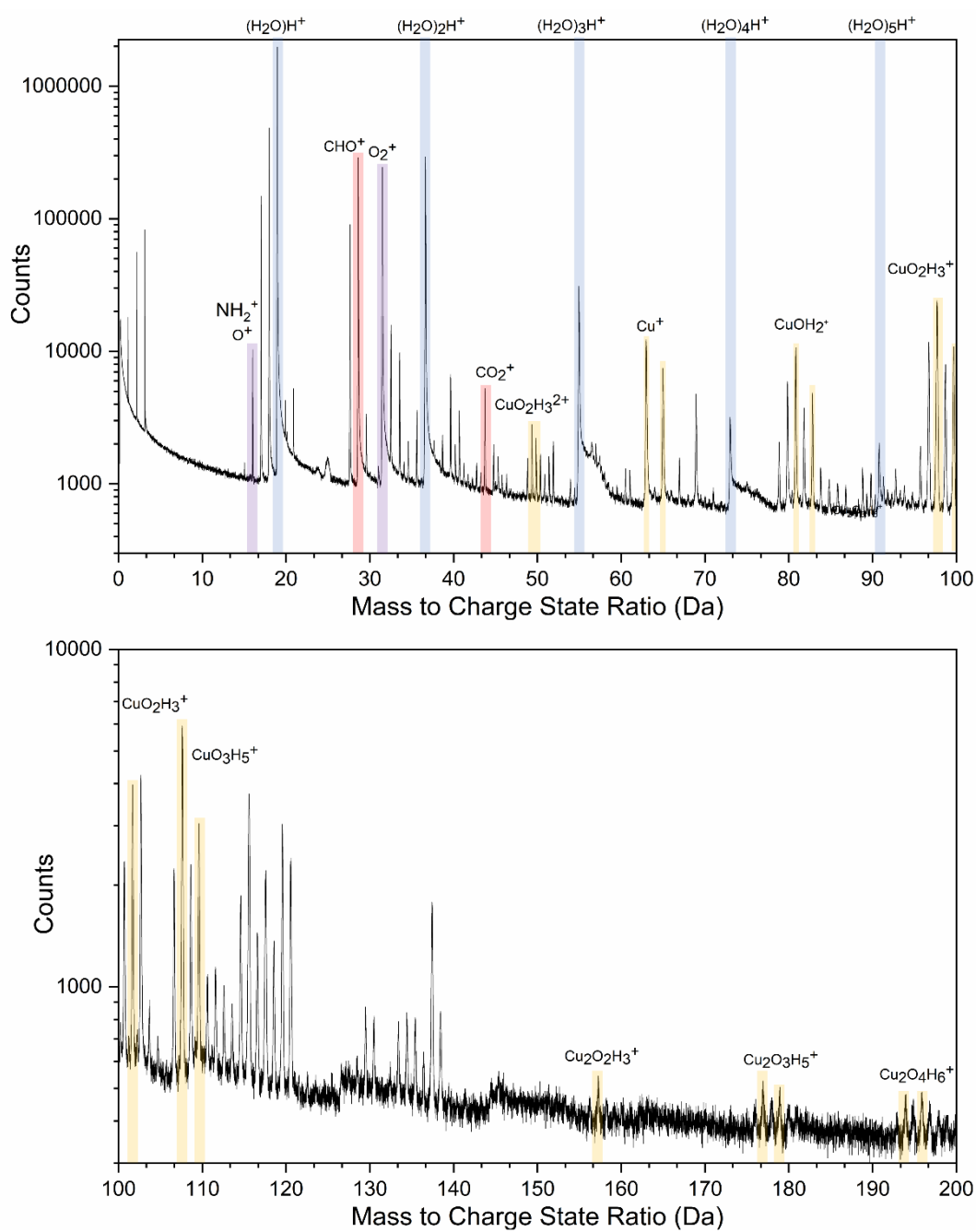

**Figure 3.** Mass spectra subset for arginine on CuZn at 140 pJ LPE (35.6 million ions, CIR: 17.2). Both sections are plotted according to magnitude of largest peak, from 300 to 1 million counts on the top, from 300 to 10000 counts on the bottom.

A dataset containing over 140 million ions was acquired from a solution containing arginine, $C_6H_{14}N_4O_2$ with a mass of 174 Da, in Type 1 ultrapure DI water. A representative mass spectrum of the arginine solution is shown in Figure 3, with the chemical structure of arginine inset at the upper right. The laser pulse energy was changed over the course of the analysis, with 40M ions acquired at 40 pJ, then the energy was increased by 20 pJ every 5 million ions, until 140 pJ where 40 million ions were collected.

Since no significant changes were seen, that data is not shown. As shown in **SI 5**, protonated water cluster ions up to n=5 are present at 140 pJ, along with a variety of signals stemming from the metallic substrate and organic fragments, with the most abundant labelled and identified. The chemical structure of arginine is shown in the inset in the upper right. There are low abundance higher mass clusters at mass-to-charge 173 Da – 180 Da and 193 – 199 Da respectively, which could be interpreted as either $Cu_2O_3H_x$ and $Cu_2O_4H_x$ or intact arginine and protonated arginine with one water of hydration. However, based on the results for the dataset in **SI 8** showing the presence of $Cu_2$-cluster hydroxides at those mass-to-charge-state ratios, they are more likely to be $Cu_2^+$ containing species. In order to try to identify changes in fragmentation of arginine or other ions caused by changes in the electric field either as the laser pulse energy is changes or simply across the field-of-view, the data was segmented into two sections of 40M ions acquired at 40 pJ and 140 pJ laser pulse energy. In these two, five cylindrical regions-of-interest (ROIs) were distributed in a grid pattern, and the calculated CIR for each ROI is shown in **SI 9**.

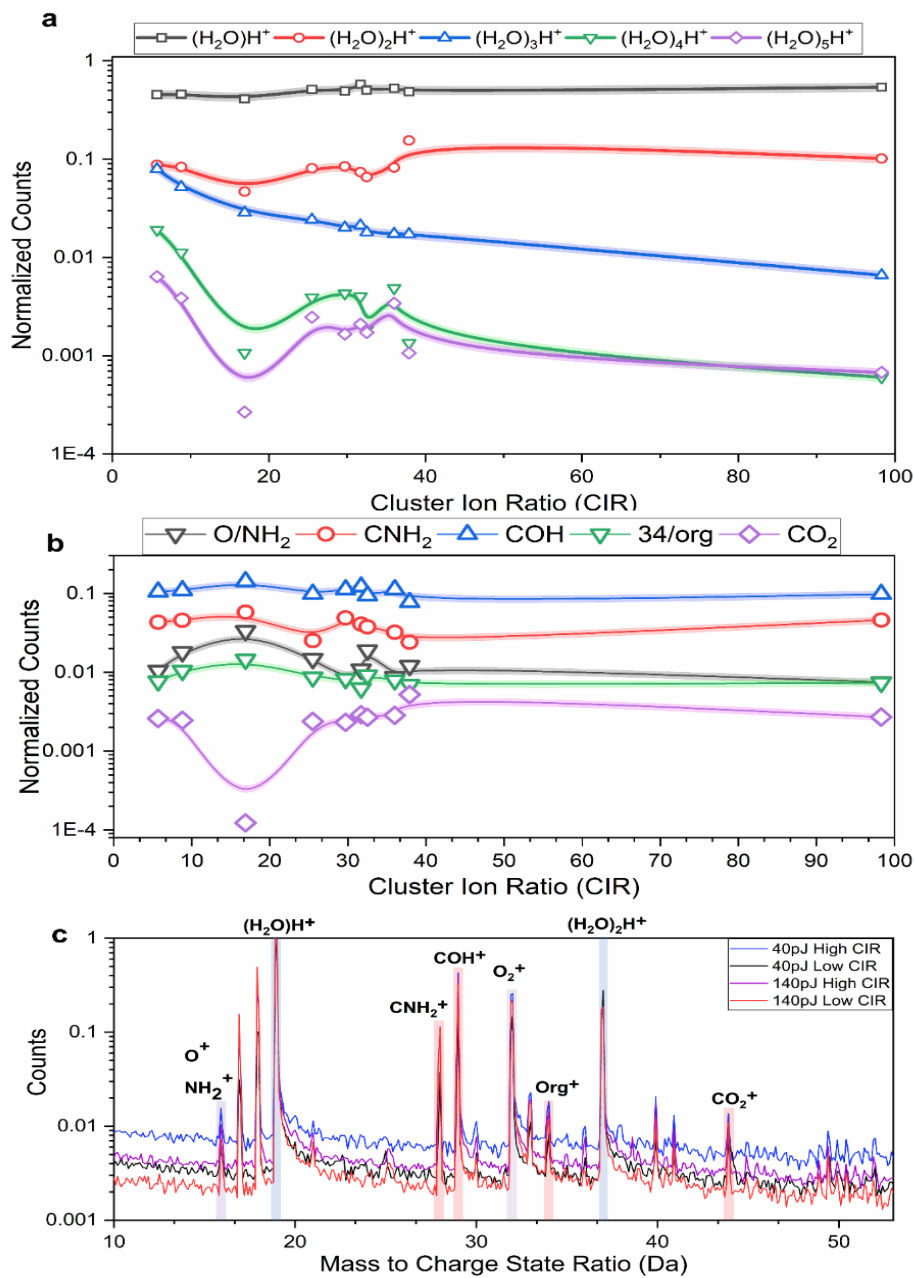

**Figure 4**. Ion abundance derived from arginine dataset. (a) Protonated water cluster abundance plotted as a function of CIR. (b) Abundance of selected organic ions as a function of CIR, with mass-to-charge state ratio less than 60 Da and known Cu species in a double-charged mass-to-charge state ratio are excluded. (c) Mass spectrum for 15 – 55 Da region.

These cylindrical ROIs in the two large sections of the dataset at different laser pulse energies (LPE) containing an amino acid provide an ideal environment to evaluate the influence of the CIR on the data. **Figure 4(a)** plots the water cluster abundance as a function of CIR, whereas **Figure 4(b)** shows the change in abundance versus CIR for various organic ions having mass-to-charge less than 60 Da, which can be attributed to organics. Above, there are too many interferences with Cu and Cu-containing ionic species to draw any conclusions. Finally, **Figure 4(c)** plots the mass spectra in the 15 – 55 Da region.

There are a number of important observations, which can be concluded from these plots. First, the higher order clusters $n$=2 – 5 demonstrate the expected decline in abundance with increasing CIR, in contrast the abundance of $(H_2O)H^+$ slightly increases with higher CIR. Similar CIRs have been observed for both voltage pulsing and laser pulsing [61]. The deviation in behaviour for $(H_2O)_2H^+$ compared to the other clusters was previously noted [34,49] and will require further investigations but could be due to . It should be noted that the peaks for the water clusters in mass spectra obtained from a region-of-interest closest to where the laser impacts the specimen had significant thermal tails, **Figure 4(c)**, which may also affect the quantification in part.

Second, the behaviour of the organic fragments varies substantially when compared with the water clusters. Note that the peak at mass-to-charge 16 Da may be $O^+$ or $NH_2^+$. The peak at mass-to-charge 34 Da is neither Cu-based nor water-based, and there is no sulphur in the solution or the considered amino-acid. It is hence presumably organic in nature, and could be $C_5H_8^{2+}$, i.e. corresponding to the complete C-backbone of the arginine. Even if rare, similar doubly-charged carbonium ions have previously been reported [85–87]. Finally, as seen in **Figure 4(c)**, the variable level of background can partially contribute to these trends. Since there was no effective way to remove the background consistently, it was not attempted. Overall, the evolution of the abundance appears to vary substantially by ion type, and does not show as pronounced a trend as the protonated water signals/clusters.

## 2.4 Cysteine

**Figure 5** plots the mass spectrum obtained from APT analysis of a solution of the sulphur-containing amino acid cysteine, $C_3H_7NO_2S$, with a mass of 121 Da, drop cast onto NPG and plunge frozen. The dataset is notable for the higher CIR of 73.2, which explains why the highest peak is $H_2O^+$ not $(H_2O)H^+$, and the higher fraction of the low mass organic fragment ions. This is likely due to the proximity of the interface with the metallic substrate, which also explains the high amount of Au- and Ag- related ions in the spectrum, marked in golden and grey, respectively. There is a relatively large peak at mass-to-charge 35 Da, along with peaks at 32 and 34, which may be related to S.

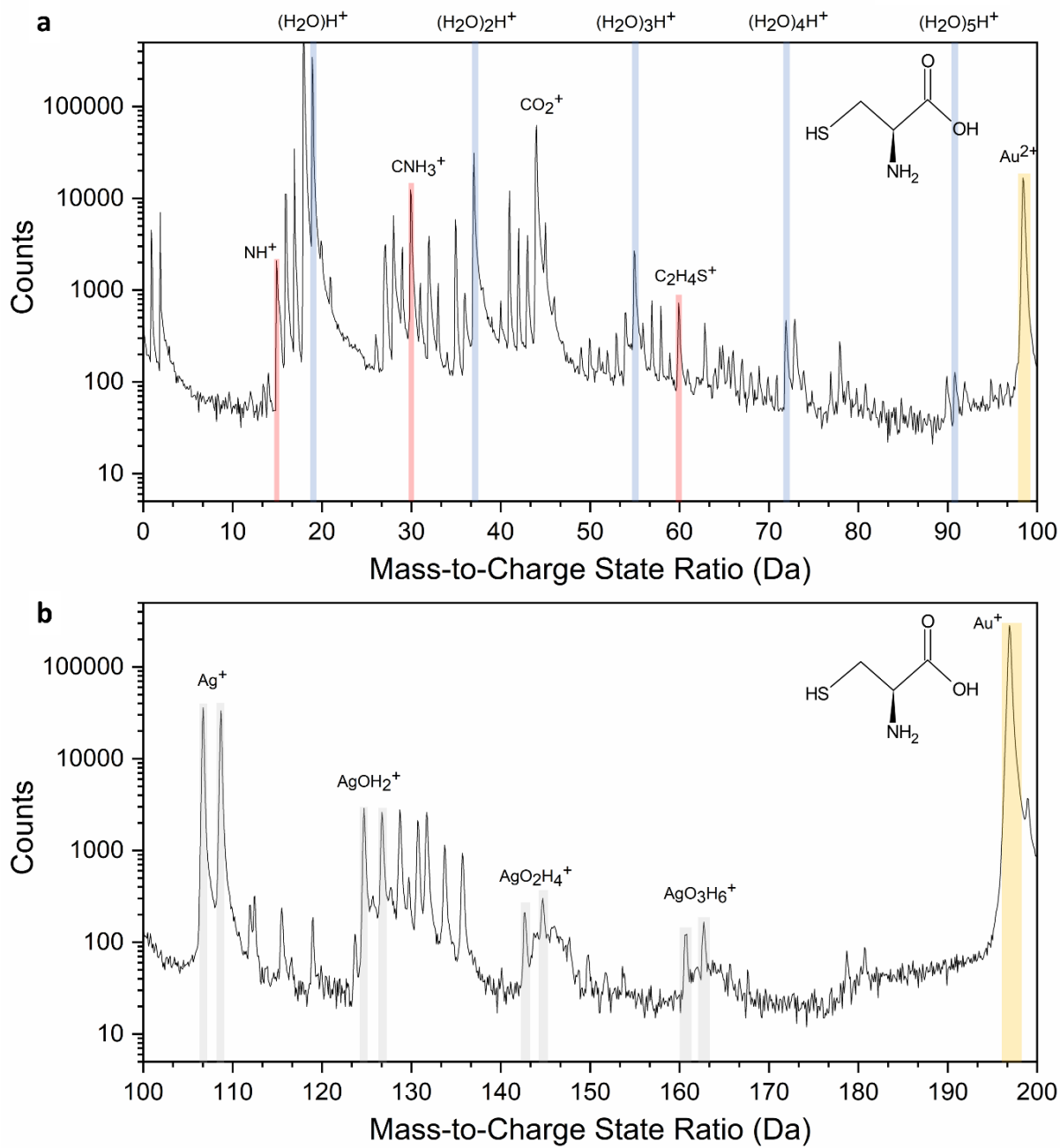

**Figure 5.** Mass spectrum cysteine drop cast onto NPG (4.5 million ions, CIR 73.2) (a) 0 - 100 Da and (b) 100 – 200 Da. The structure of cysteine is inset in the upper right. The sulphur moiety is located at mass-to-charge 32 Da, although partially obscured by the $O_2^+$ ion contribution, and also at mass-to-charge 33 Da in the form of $SH^+$, also partially masked by the $O_2^+$ thermal tail. Note the occurrence of Ag-water clusters in (b).

To aid with identification of S-containing peaks, pure sulphur was cryogenically prepared via liftout and coated with Cr (Schwarz, et al., 2024). The corresponding mass spectrum is plotted in **SI 10(a)**. $^{32}S^+$ is dominant and visible above background, but $^{34}S^+$ is not; the other, notable mass-to-charge peak at 34 Da is $SH_2^+$ with secondary low-abundance $SH^+$ and $SH_3^+$ peaks as 33 and 35 Da respectively. This can be determined since the isotopic ratios do not match for $^{32}$S and $^{34}$S. However, arginine does not contain S but the data shows peaks at mass-to-charge 33 Da and 34 Da in the enlarged mass spectrum in **SI 10(b)**, therefore those peaks are organic and not S-derived.

There is a relatively large peak at mass-to-charge 35 Da. Since all amino acid solutions were prepared from hydrochloride salts, when comparing arginine and cysteine datasets as in **SI 10(b)** and **SI 10(c)** respectively, a peak at mass-to-charge 35 Da is present in both. While Cl$^+$ or HCl$^+$ were not observed in APT study of an aqueous solutions containing NaCl [34], those ions were observed in minerals and metals containing fluid inclusions [88,89] and in other specimens like PVD films [90], corroded steel [91], and other geologic specimens [92]. It is well known that the detection of Cl ions is difficult due to the high ionisation energy required, but as described above, it was possible to detect Cl$^+$ in various systems. Base on the use of the hydrochloride salts we assume that the signal at 35 Da is Cl$^+$, but the disappearance of the isotope at 37 Da in the (H$_2$O)$_2$H$^+$ peak makes it difficult to unambiguously define it as Cl. Therefore the peak at mass-to-charge 35 Da is assigned to Cl$^+$.

## 3  Discussion and perspective

### 3.1  Substrate and specimen preparation

Metallic ions in solution are functionally impurities more likely found when the final tip of the specimen is in closer proximity to the metallic substrate. As discussed in Ref. [35,56,93], Cu-containing molecular or cluster ions are commonly observed in APT of wet samples, whether created from chemically dealloyed CuMn, CuZn, or dissolved Cu incidental to graphene encapsulation. Likewise, Zn in the presence of O has been reported to evaporate in cluster form in APT [94]. In the context of organic compounds, free or solvated amino acids, particularly arginine and lysine, are known to form stable divalent metal (II) complexes with either copper or zinc [95–97]. This has been validated for most amino acids as well, although this does not seem to involve the terminal amine group for lysine [98,99] and likely not for arginine. Most textbooks state that only five amino acids are zwitterionic, which have variable side groups, two of which are studied here: lysine and arginine. However, that assumes that the amino acids are present in a protein and both the carboxyl and amine ends are bonded together. Free amino acids in solution with both unterminated carboxyl and amine groups are therefore zwitterionic, and that may effect the resulting evaporation and fragmentation. In the mass spectrum in **Figure 3**, ions which likely correspond to CuCN$^+$ complexed with H$_2$O are found. These ions might form during the field evaporation process or with any amino acid in solution with a free amine group, and in any case can be related to the backbone amine group rather than the terminal ω-amine group in lysine or arginine's side chain [99].

This behaviour would definitely change the resultant mass spectrum and lead to the detection of those bonded amines as heavier molecular ions, but also change the relative abundances of other amino acid fragments. These higher order fragments in the analysis and reconstruction of amino acids, because the spatial proximity of ions such as $Ag(CO)^+$, etc., to other nitrogen-enriched fragments is necessary to determine the identity of specific molecules. Additionally, for this work, the samples were manually plunged into a liquid nitrogen bath. The freezing rate was not sufficient to vitrify the samples, leading to the non-aqueous components of the solution to segregate domain-like structures in the reconstructed APT data, particularly for the analysis of arginine in **SI 9**. These domains exhibit a higher CIR, and a higher relative amount of multiple hits.

The further refinement and wider deployment of substrate-free FIB liftout techniques to prepare specimens from bulk frozen liquids is necessary to facilitate creating simpler mass spectra [41], along with alleviating requirement for nanoporous metal substrates. Bulk sample preparation, through e.g. high pressure freezing (HPF) that can create amorphous liquid discs over a hundred microns in thickness, will also help maintain the biomolecules in their native state. Such samples can then be transferred under cryogenic conditions in liquid nitrogen to the FIB and lifted out. This preparation method would avoid the need for nanoporous metals altogether. In the alternative or additionally, better preparation methods for nanoporous metals to completely remove the sacrificial metal (either Ag or Zn) would facilitate mass spectral analysis.

## 3.2  CIR & electric field

The variations in specimen shape between experiments and during a single experiment can make it challenging to obtain reproducible data, and makes the LPE a poor descriptor of the electric field condition, as shown in **Figure 1**. The variation of the CIR across the field-of-view, as shown in **SI 9**, could partially have been expected based on the proximity to the location where the laser illuminates the specimen. The increased temperature there inherently has a lower electrical field and the tip shape becomes flatter on that side, as reflected in the lower CIR there. There are three levels of hierarchy and structure which are relevant here. At the mesoscale, as noted above, tip shape is influenced by laser direction. At the multi-nanometre scale, the slow freezing process did not vitrify the solution, and so there are organic rich regions. Functionally, these behave like precipitates in a matrix and possess higher evaporation fields, which create local topography and evaporate at different rates than the water matrix [100,101]. At an atomic or local level, the evaporation of the water matrix depends on cluster ion size [36]. Therefore, some of the observed CIR variations are related to local composition variations because of the organic rich regions found randomly within each sample cylinder, as well as to localised reshaping of the specimen under the combined influence of the laser pulsing and the intense electric field [36].

The variations in specimen shape between experiments and during a single experiment can make it challenging to obtain reproducible data[102], and makes the laser energy a poor descriptor to describe the electric field condition, as shown in **Figure 1**. As had been indicated in work on e.g. oxides [63,103] charge state ratios overall are a good electrical field strength proxy. The variation of the CIR across the field-of-view (FoV), as shown in Figure **SI 9**, could partially have been expected based on the proximity to the location where the laser illuminates the specimen. The increased temperature in this region causes a specimen shape evolution to locally increase the curvature and leads to a lower electrical field on that specific side of the specimen [69], which is reflected by the lower CIR.

There are three levels of hierarchy and structure which are relevant here. At the mesoscale, as noted above, tip shape can be influenced by laser incidence direction. At the multi-nanometre scale, the slow freezing process did not vitrify the solution, and therefore are organic rich regions are formed. Functionally, these regions behave like agglomerations or precipitates in a matrix and possess higher evaporation fields, if the evaporation field strength of the organic material is assumed to be higher than the water matrix itself. Therefore,create local topography and evaporate at different rates than the water matrix, which leads to a distortion of the trajectories of the evaporated ions, local magnifications and a drop in spatial resolution [100,104]. At an atomic or local level, the evaporation of the water matrix depends on protonated water cluster ion size [36]. It has been shown that water clusters are pulled out of the surface of the tip and field evaporate [50], however, nothing is known about the evaporation behaviour of biological molecules such as amino acids in a water matrix, which can lead to an inhomogeneous sample surface and thus to an uneven electric field distribution. Therefore, some of the observed CIR variations can be related to the surface structure and the related local composition variations due to the formation of the organic rich regions, which are found randomly within each sample cylinder, as well as to localised reshaping of the specimen under the combined influence of the laser pulsing and the intense electric field [36].

The overall implications of the CIR on the post-evaporation dissociation and fragmentation behaviour of the individual molecules is yet to be quantified, but **Figure 4(b)** suggests a variability. This introduces unnecessary variability and potential confounders into the data, including across the field of view. There are many unknowns as to whether the fragmentation depends on the molecule's orientation relative to the field evaporating surface [105], but it likely depends also to their local chemical environment, i.e. presence of spurious metals, that affect the local intensity of the electrostatic field. To some extent, this explains the APT data of proteinaceous structures, e.g. an IgG antibody embedded in a silica matrix [106] showing extreme fragmentation, such that only $CNH_3^+$ and $CO_2^+$ ions were recovered in high proportion.

### 3.3 Fragmentation and amino acid identification

Our data shows that it may be possible, even if challenging, to control the distribution of the collected fragments from amino acids, as in **Figure 4**. At lower CIR, generally heavier (larger) organic fragments are detected, i.e. less subject to fragmentation, but protonated water cluster ions up to higher orders will be measured, which can hinder the detection of heavier organic fragments. At higher CIR, lighter (smaller) fragments are detected, along with lower order protonated water cluster ions, meaning higher order protonated water clusters are suppressed to background albeit with larger molecular ions fragments. Smaller fragments on the one hand would improve the spatial resolution, but a multiple of small fragments can lead to a more complex peak identification due to the high number of overlapping signals. On the other hand, the detection of larger fragments up to whole protonated molecules allows the localisation and identification of whole organic molecules and their distribution in the 3D volume [37]. There is hence a need to balance these two aspects in the optimisation of the analysis conditions.

Practically, the identification of amino acids will likely be based on partial sets of fragments of the initial molecule. The lack of pronounced tracks or "hot spots" in the correlation histograms indicates that the field fragmentation takes place at the specimen's surface during the field evaporation process itself [105]. It appears, that the post-evaporation dissociation mechanism favours the removal of a hydroxyl group followed by the second oxygen (or the carboxylic acid group entirely), meaning that identification of the amino acid will have to be based on molecular ions lacking oxygen. Those could also be complexed with metal ions which are in solution as well. These considerations have several implications for the identification of individual amino acids within proteins, since practically the only difference in each amino acid is the side group. Identification of hydroxyl or carboxyl groups will likely be one strategy for attempting to identify individual amino acids within a protein.

One other significant factor for reconstructing water and biomolecules, which applies to oxides as well, is the consideration that both oxygen and nitrogen atoms tend to form neutral $O_2$ and $N_2$ species that are typically not detected, though their presence can be inferred from correlated evaporation histograms [64]. Several pathways typically exist for oxygen to dissociate and form neutral species from oxides [73]. Likewise, the loss of nitrogen as neutrals has posed substantial problems in analysing III-V nitrides in the APT [107–109]. This loss mechanism is additional to the normal ion loss in a reflectron system and loss of spatial accuracy in a reflectron system.

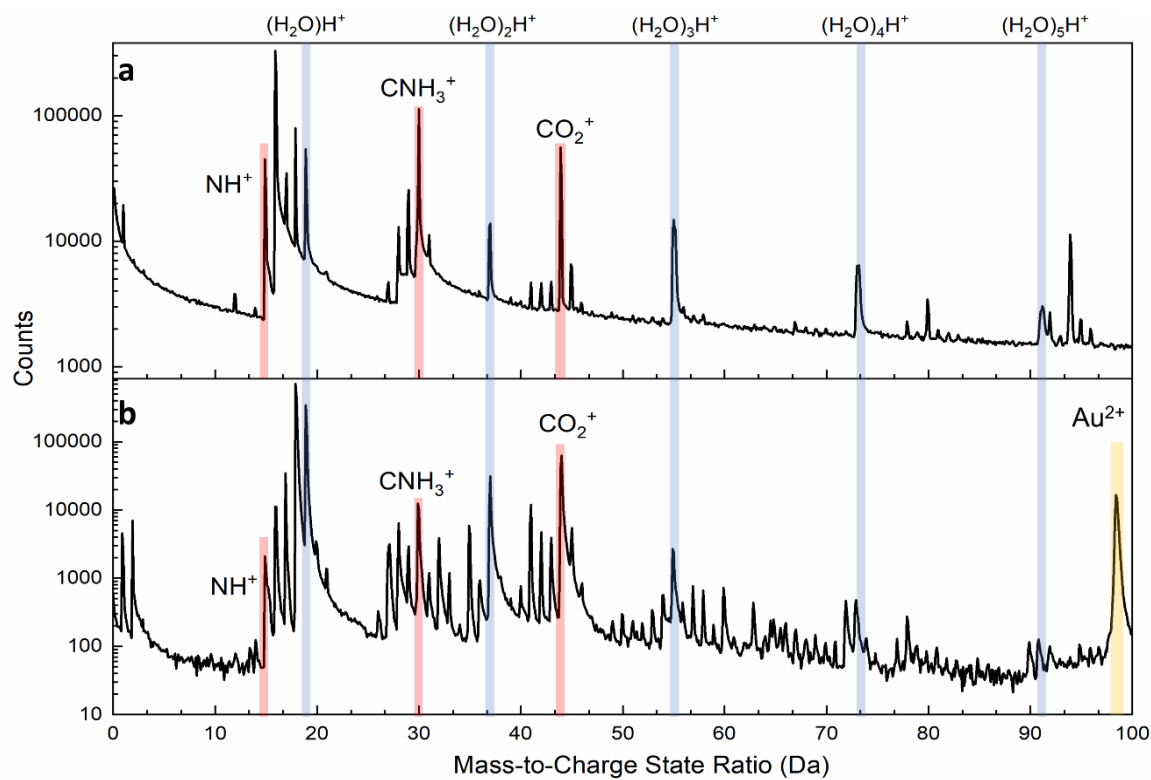

**Figure 6.** Comparison of lysine dataset (CIR 3.82) versus cysteine (CIR 73.2) over same mass spectral range, limited range of 0 - 100 Da because of the metal ion overlaps starting at mass-to-charge 98.5 Da for $Au^{2+}$

**Figure 6** plots the mass spectra for lysine in water on NPG, although with no visible metallic ions, versus cysteine in water on NPG with a CIR of 73.22. This CIR differential is substantial, although interestingly water clusters up to n=5 appear visible in both cases. Overall, there are differences in the spectra, although the most common ion set for amino acids or proteins in the APT literature are still the most prominent, e.g. $CO_2^+$ and $CNH_3^+$ / $COH^+$ [43,44,52,81,106]. Whether these differences are going to be exploitable to differentiate between amino acids will be the focus of future work. Yet, this result provides hope that amino acids can be differentiated by their side chain ionization patterns, which is encouraging for the future. However, in order to firmly assign peak identities, ab initio and/or molecular dynamic simulations will be necessary, similar to e.g. Gu et al. [110] but adjusted for the electric field conditions in APT [73]. As an example, the fragmentation behavior of glucose ring in a pure water solution under a high electric field in APT was facilitated by DFT simulations, as in Ref. [37].

## 3.4 Perspectives on future work

APT is often considered to offer calibration-free quantitative analysis of materials. Although this may be (nearly) true for some metallic systems, this view likely needs to change, considering the field-dependence of the measurement accuracy and possible species-specific losses [64,71,73,109,111]. To create fragmentation profiles for each amino acid, standardized reference spectra for each amino acid will likely need to be acquired beyond the cysteine and lysine presented here in this work. More specifically, the reference spectra should be acquired under similar local electrical fields, which the CIR can help guide and compare different datasets in future with each other. This work demonstrates that the CIR effects peak heights in the mass spectra for two amino acids, at least via control of water clusters and controlling the overall background level. The fragmentation of lysine versus arginine as shown above illustrates there are substantial differences. These spectral libraries can likely then be used to identify individual fragments and inform further amino acid identification, through the use of machine-learning algorithm, some of which already exist for APT but will need dedicated adaptation [112].

There is an elephant in the room, which that electron imaging and Ga- or Xe-based focused ion beams used for preparing specimens have the potential to induce substantial damage[113–116], including in metals[117,118]. For metallic materials, using cryogenic temperature during FIB-based APT specimen preparation has shown substantial benefits[119,120], however, for frozen liquids little has been documented, and there are debates in the cryo-TEM community as to the [116,121]. Peaks pertaining to Xe or Ga are not typically observed in our analyses, but they may also be obscured by peaks pertaining to other molecular ions, and in metals, structural damage can extends tens of nanometers below where implanted ions are imaged[118]. What part of the obtained results are affected by beam damage remains unknown, even if we expect that this will mostly affect the early stages of the APT analysis that are typically discarded from the analysis.

With regards to the complexity of the analysed solution, a caveat is that for combinations of amino acids, the peptide bond will change the fragmentation and resulting in different spectra. Dipeptides, for example lysine coupled with lysine, have been analysed when dried on a small ball of carbon nanotube (CNT) mesh using APT, where higher mass fragments corresponding to the combination peptide were observed [42,44]. Those results demonstrate that higher mass molecular ions, which represent protonated dipeptides, can be detected, as well as dipeptide clusters combined with other fragments. Analysis of similar dipeptides in aqueous solution should be tried to see whether comparable results can be achieved.

One could assume that the fragments from a single amino acid molecule will be detected in close proximity spatially and temporally – i.e. within a single pulse or over the next few pulses as had been shown previously for other material systems [122,123]. However, when the large arginine dataset was thoroughly analysed, higher order multiple hits present in the filtered data (e.g. three or more) look like the overall dataset, e.g. the evaporation is temporally uncorrelated. The need for high spatial resolution and high detection efficiency will likely favour the use of straight flight path APT systems, whereas most commercially available and active APT systems are reflectron-based, with approx. 50 % ion loss, vs. less than 20% for a straight flight path instrument. Then new targeted data processing tools for investigating spatial and temporal correlations between events, not unlike what has been used in the recent past to look for correlated events, will have to be developed [76,124].

Ultimately, the goal is not to analyse individual amino-acids but complex assemblies into proteins. Here, we used unlabelled Aβ1-42 fibril in solution [57], which were drop-cast and frozen by plunge freezing into LN2, transferred to the cryo-FIB for specimen preparation, and analysed in the LEAP 5000 XS. A ROI inside the NPG containing is shown in **Figure 7(a)**, with protonated water cluster ions displayed as blue spheres, organic fragments in red, and the gold isoconcentration surface indicates the boundaries of the metallic pore. The red isoconcentration surface evidences a high carbon concentration consistent with a trapped amyloid fibril located inside an NPG pore. The mass spectrum for the data within the isoconcentration surface is plotted in **Figure 7(b)**. These fibrils were selected to allow for comparison with dried $^{13}$C- and $^{15}$N-labeled fibrils deposited onto an aluminium specimen as reported previously [59]. One of the mass spectra from these experiments is plotted in **Figure 7(c)**, and it consisted of hydrocarbon fragments, e.g. $C_2H_x$ – $C_4H_x$ in the dried form. These two mass spectra show a significantly different fragmentation pattern, which highlight the challenges lying ahead.

Finally, for future work also includes further development for sample preparation techniques, such as a combination of HPF vitrification and cryo-FIB/SEM preparation techniques for liftout from organic-containing bulk aqueous specimens to avoid the requirements for nanoporous metal substrates and the agglomeration of organic rich regions; better chemical and thermal dealloying methods; development of standardized individual amino acid datasets; and machine learning methods for determining individual amino acid identities within individual proteins.

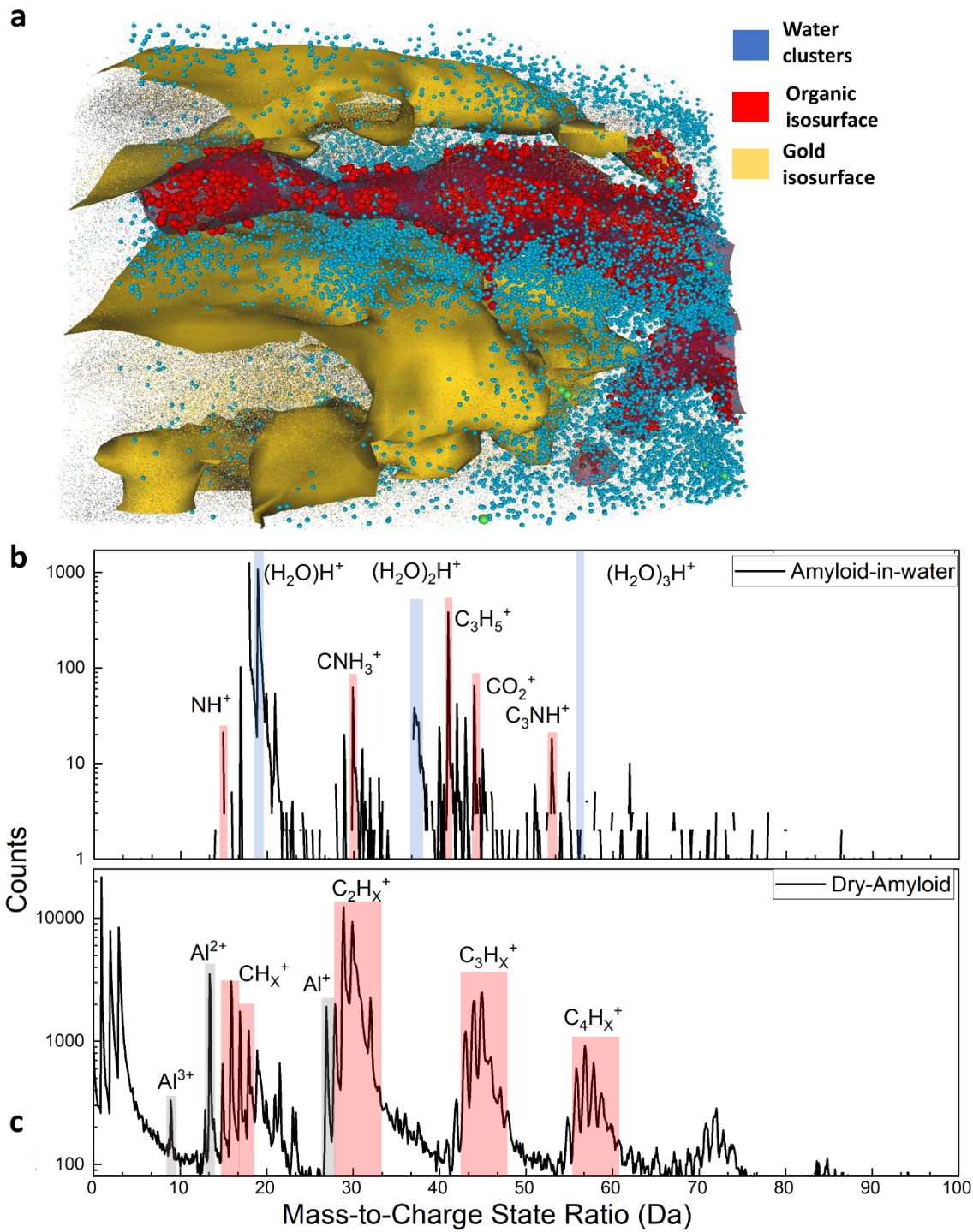

**Figure 7.** (a) Visualization of an unlabelled amyloid Aβ1-42 fibrils trapped in NPG pore, with protonated water clusters as blue spheres, and a gold isoconcentration surfaces that delinates regions containing over 70 at% (Au+Ag) and a red isodensity surface that encompasses regions containing over 1.2 C at.nm$^{-3}$. (b) mass spectrum of hydrated organic material, e.g. an amyloid fibril, trapped in NPG pore, on limited mass range to 100 Da since Au$^{2+}$ and other higher mass metal ions block visibility of any organic ions with higher mass (c) mass spectrum of original labelled Aβ1-42 amyloid fibrils dried on pre-sharpened Al tip with significant regions highlighted

# 4 Conclusion

To summarise, we have on a systematic study of amino acids dissolved in pure water by APT. We introduced the used of the CIR as a proxy for the local electric field conditions, which appears repeatable across different APT systems, independent of laser energies or acquisition mode. Through analysis of lysine and cysteine, we could demonstrate that different amino acids fragment differently and the different groups could be detected. Through analysis of the arginine dataset, we demonstrated different CIR values produce different fragmentation ratios of lower mass organic ions, which can be used in future to tune the mass spectra. We then discussed the limitations of current approaches, particularly in terms of specimen preparation and analysis, along with future development of optimized techniques for the identification of amino acid sequences within proteins and provides bounded guidance on which types of fragments are likely to be most consistently recoverable.

# Acknowledgments

The authors would like to acknowledge Uwe Tezins for his assistance with the MPIE APT facility, and Andreas Sturm for his assistance in CAD design, probe fabrication, and FIB technique development. Additionally, the authors are grateful for the assistance of Jürgen Wichert in the MPIE metals workshop for vacuum dealloying techniques for copper and the MPIE mechanical workshop for their help in designing and manufacturing different sample holders, equipment holders, and other components with CNC and EDM, particularly Rainer Lück and Tristan Wickfeld. The advice of Dr. James Douglas at Imperial College London for sample preparation and Dr. Shyam Katnagallu for data analysis is much appreciated, as is the initial help from Dr. Leigh Stephenson, now at the University of Sydney, while he was still at MPIE. Leonardo Shoji Aota si gratefully acknowledged for help and experimental support. EVW, TMS, SZ, IM and BG would like to acknowledge the DFG funding through the 2020 Leibniz Prize. BG, IM, AEZ and SHK would like to acknowledge funding through project SHINE (ERC-CoG) #771602 and the DFG through DIP Project No. 450800666.

The authors declare no conflicts of interest.

# 1 Materials and Methods

## 1.1 Materials

AuAg thin foil 100 µm thickness (Au: 25%, Ag: 75%) was obtained from Goodfellow (Goodfellow Cambridge Ltd., Huntingdon, UK). CuZn (Cu: 63%, Zn: 37%) foils and sheets of varying thickness were obtained from Metall Ehrensberger (Metall Ehrensberger GbR, Teublitz, Germany). CuZn foil 100 µm was also sourced from LLT Applikation (LLT Applikation, Ilmenau, Germany). AuAg and CuZn grids were laser cut by LLT Applikation according to own previously developed designs [56]. Bulk chromium flakes (99,9% purity) were obtained from the MPIE synthesis and processing lab.

Nitric acid (69%), hydrochloric acid (HCl) (37%), and Type 1 ultrapure deionized water (DI) were obtained from Merck (Merck KGaA, Darmstadt Germany). Lysine hydrochloride (HCl), arginine HCl, and cysteine HCl were obtained from Alfa Aesar (Alfa Aesar, Lancashire, UK). Type 1 ultrapure DI water was additionally obtained from an in-house DI water plant. Fibrils derived from recombinantly produced amyloid-beta 1-42 (Aβ1-42), either in uniformly $^{13}$C, $^{15}$N labelled state or in nonlabelled state with C and N in natural abundance , were prepared and handled as described [57–59].

Coupons with 22- and 36-post silicon microtips (Cameca FT-22 and -36, Cameca Instruments, Madison, WI, USA) were used for mounting lift-out specimens. Coupons were then mounted on Cameca Copper Spring Clip Microtip Stubs, to be placed in Cameca specimen pucks ("clips"), equipped with the appropriate polyether ethyl ketone (PEEK) plastic isolating collar ("cryo-puck") in either single post or dual-post versions, as appropriate. So-called Felfer holders (e.g. [60]) were either made internally by the MPIE mechanical workshop (original design) or a slightly newer version (Specimen Holder Mk I, Microscopy Supplies Australia, Macquarie Park, Australia) were used.

## 1.2 Methods

Bulk nanoporous gold (NPG) was obtained from the AuAg foils, 100 µm thickness, punched into 3 mm disc shapes using a TEM punch. These discs were immersed in $HNO_3$ (65%) for ten minutes and removed, then rinsed thoroughly in Type 1 ultrapure DI water multiple times and left to soak for eight hours typically, as described in [34]. Grids were prepared using the design from Ref. [56] and dealloyed as above.

CuZn foil, 100 µm thickness, was punched into 3 mm disc shapes. Some 100 µm CuZn TEM half-grids were laser-cut by LLT Applikation, using the design shown in Ref. [56]. CuZn sheets 0.4 mm thickness were pre-cut into crowns (e.g. a vertically self-aligned piece with 6 individual sharpened spikes on one end), as described in in Ref. [56], along with the metal heat treatment protocols to remove any stress or damage during manufacture. Nanoporous Cu was created by soaking for 4 hours in HCl, followed by thorough rinsing with Type 1 ultrapure DI water, as described by Woods et al. [56]. Substrates were immediately dried using argon and stored under nitrogen until used. Amino acid solutions were prepared with Type 1 ultrapure DI water to 0.1 M concentration.

The cryogenic sample preparation, fabrication, transfer, and analysis infrastructure at MPIE, including FIBs, APTs, cryo-transfer systems, etc., are described in previous work [32,41] and is summarized here. FIB/TEM half-grids were mounted in Felfer holders. Those were then mounted in Cameca cryo-pucks either inside or outside the nitrogen-filled Sylatech GB-1200E glovebox (Sylatech GmbH, Walzbeuchtal, Germany) ("glovebox"). as appropriate. If / when applicable, tips were immersed in liquid for different amounts of time to fill pores or the solution was drop cast as a 2 μl droplet on the surface and blotted to remove excess liquid. When applicable, such NPG discs were mounted on a Cameca copper coupon holder, under the copper spring, immediately before use. Samples were directly plunge frozen into liquid nitrogen and loaded in a Ferrovac D-100 ultrahigh vacuum cryogenic transfer module (UHVCTM) (Ferrovac GmbH, Zurich, Switzerland) cooled to -190°C using liquid nitrogen (LN2), hereafter referred to as "suitcase."

After transfer, samples for APT analysis were made using either a Thermo-Fisher Helios G3 Xe plasma FIB (PFIB) (Thermo-Fisher Scientific, Billerica, MD, USA) equipped with a Gatan C1001 cryo-stage cooled to -190°C (Gatan Inc., Pleasanton, CA, USA). The annular milling procedure for APT specimen sharpening and the specific beam currents used are the same as in Ref. [34]. Alternatively, the cryogenic lift-out procedure for specimen preparation described in Ref. [61] was used on a Thermo-Fisher Helios G5 CX Ga FIB / SEM equipped with a freely rotating Aquilos cryo-stage cooled to -190°C and a cooled cryo-EZ-Lift manipulator. The specimen was mounted onto an individual post using redeposition welding onto a commercial Si-coupons in dual-post pucks, and sharpened to an end diameter less than 100 nm. At that point, the specimens were ready for APT analysis. Hereafter they are referred to "tips."

The pucks were transferred back to the suitcase and moved to the APT for analysis. Two Cameca APT systems were used, each equipped with a 355 nm UV laser – a LEAP 5000 XS straight flight path and a LEAP 5000 XR reflectron system. Details of the laser powers and other relevant run parameters discussed in the next section are provided in supplementary data table **T1**.

# Supplemental Information

**SI 1.** Correlated evaporation histogram lysine in water with no metal ions present

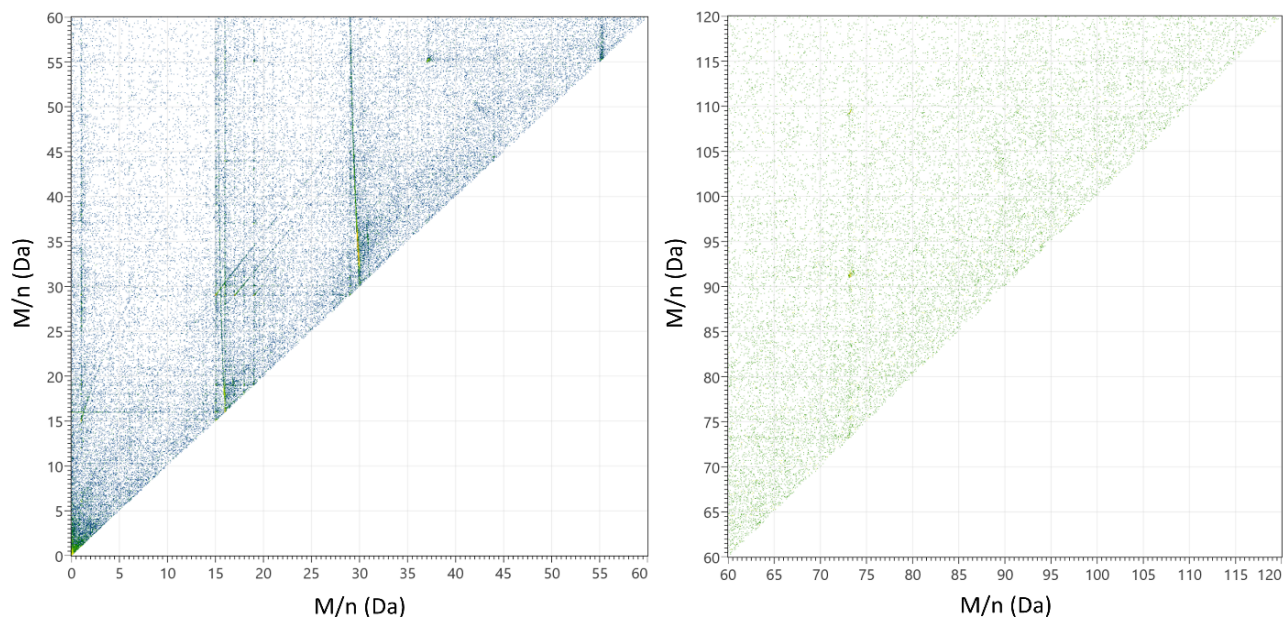

**SI 2.** Correlated evaporation histogram lysine in water on AgAu

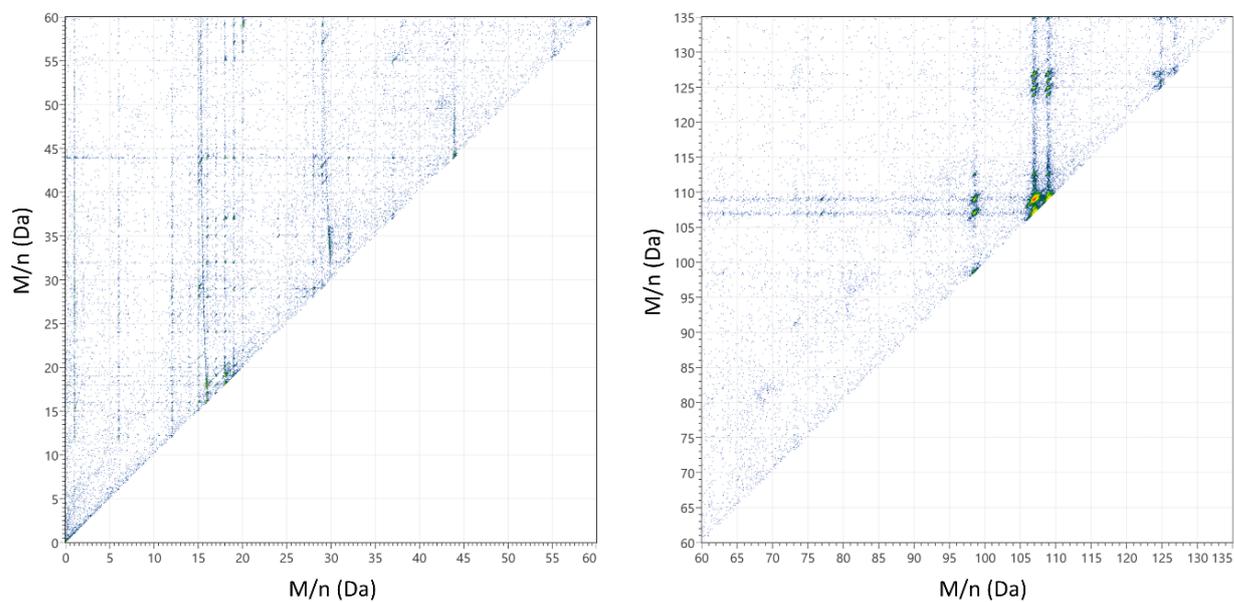

**SI 3.** Enlarged and labelled mass spectrum for lysine in water mass spectrum for Ag-rich mass spectral range

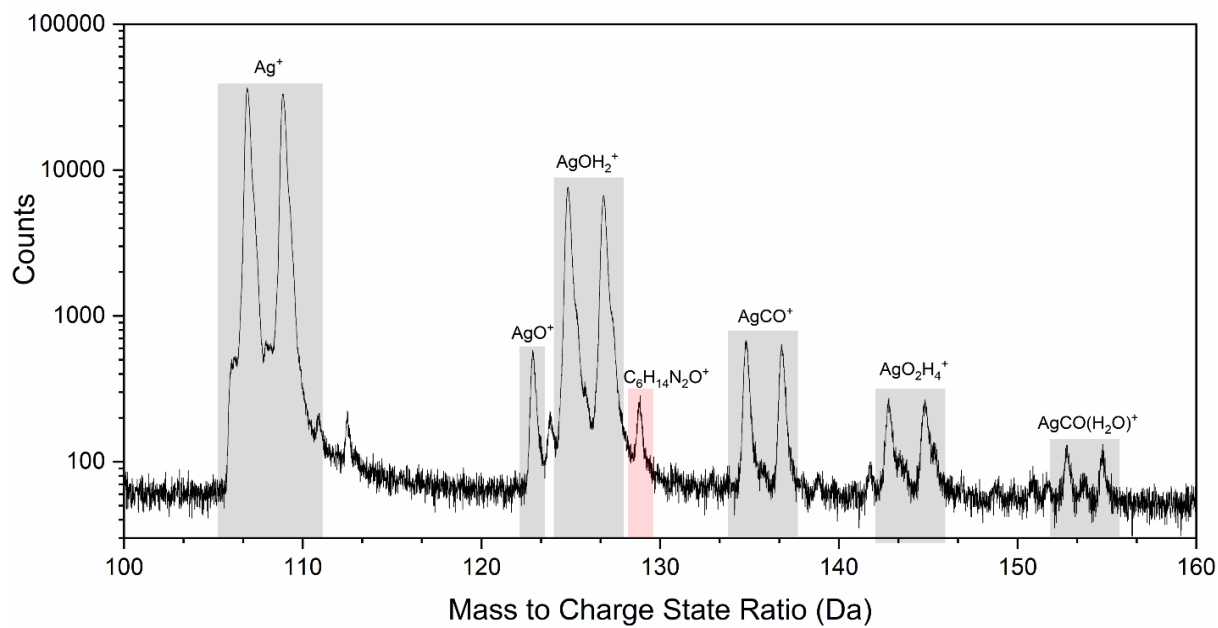

**SI 4.** Enlarged mass spectrum for lysine in water on AgAu with labelled Au water clusters and carbonates for primary Au-rich mass spectral range

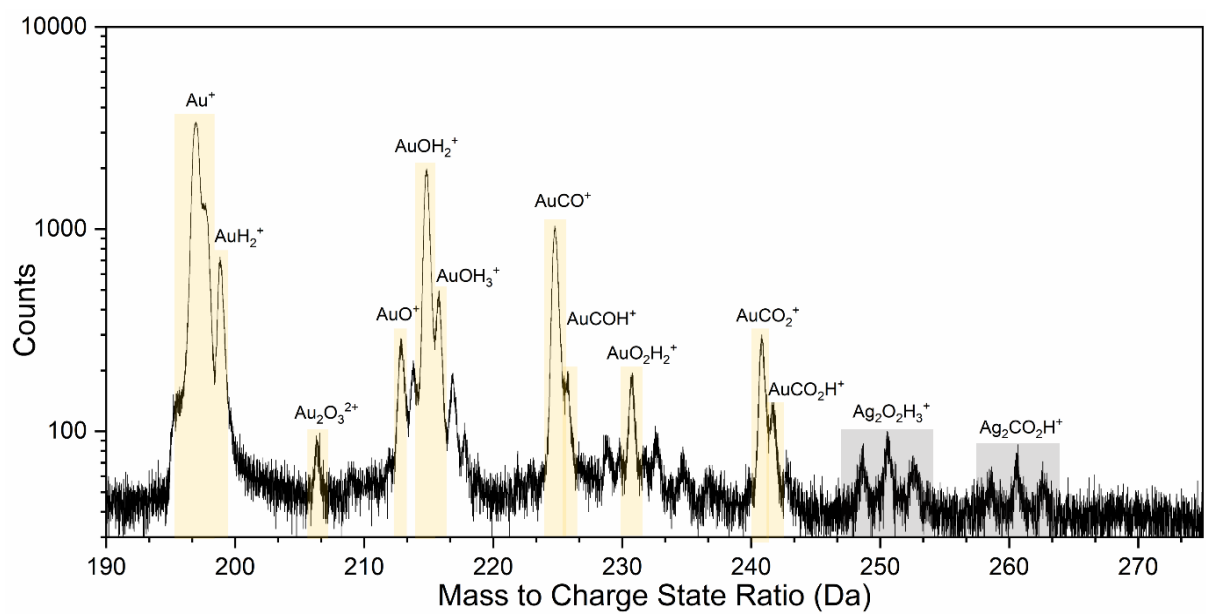

**SI 5.** Enlarged mass spectrum for lysine in water on CuZn with labelled Cu hydrates and oxides in Cu-rich mass spectral range

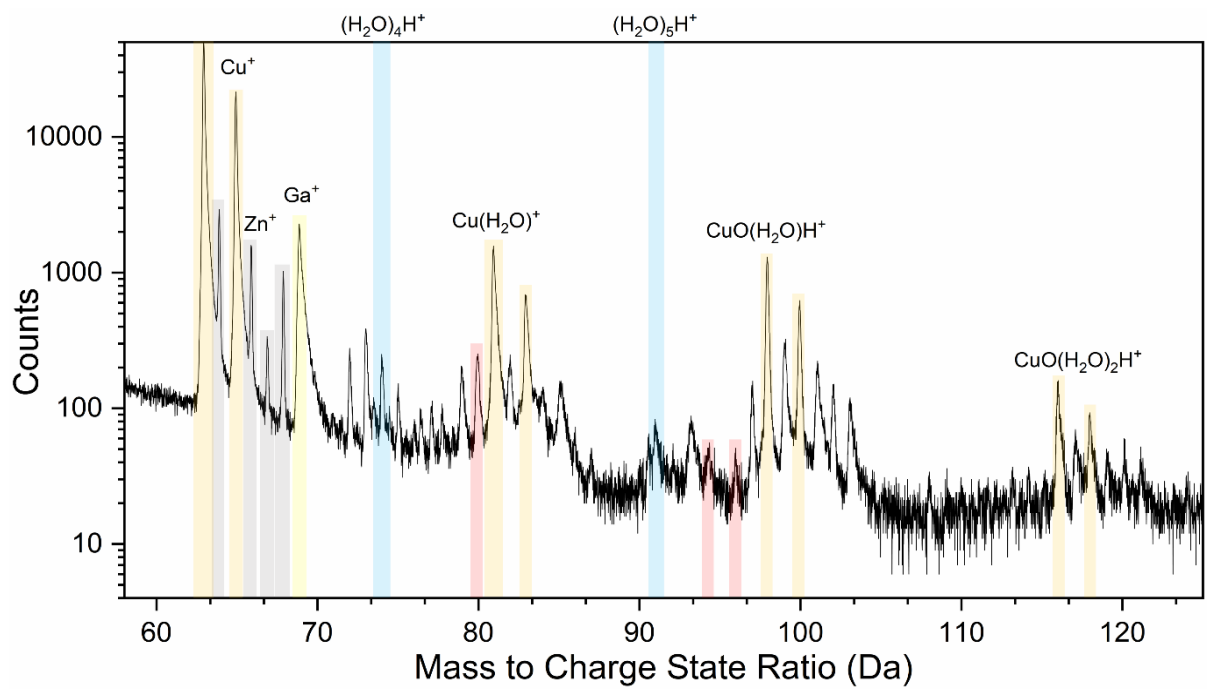

**SI 6.** Correlated event histogram lysine in water on CuZn

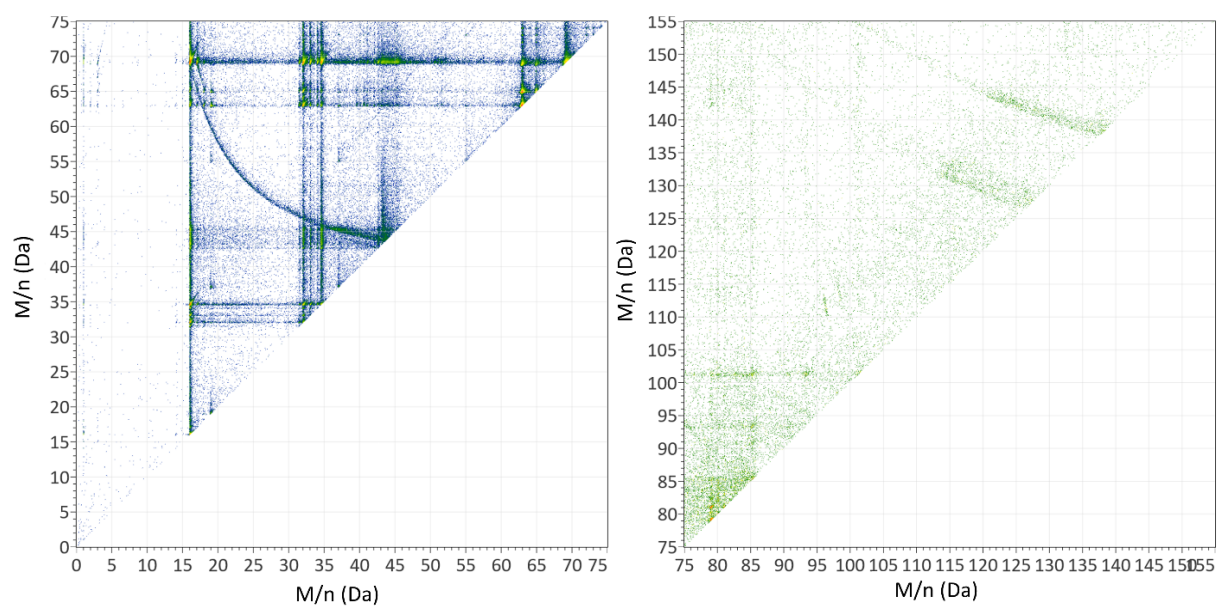

For the specific case of the left graph in **SI 6**, conservation of mass and charge requires a parent ion having mass-to-charge 44 Da in a 2+ charge state, which decomposed into two product ions, each in 1+ charge state. One of those was at mass-to-charge 69 Da, possibly either $C_4H_7N^+$ or $C_3H_3NO^+$, and one at mass-to-charge 16 Da, likely $O^+$, on the left, the parent ion would likely be $C_4H_{10}NO^{2+}$ or $C_3H_6NO_2^{2+}$ having mass 88 but mass-to-charge 44 in this case [42,44,80]. Possibly the $NO_2^+$ containing ion could become resonance stabilized, although MD simulations for ion stability would be required to fully answer this question. A closer examination of the histogram reveals two curved tracks in **SI 7** proceeding to mass-to-charge 69 Da on the y-axis. The first several million ions of the dataset were excluded to avoid this possibility. One is located at mass-to-charge 16 Da ($O^+$) and the other at mass-to-charge 17 Da ($OH^+$).

In classical electrospray MS, a water would typically fragment from a parent ion; in this case the high electrical field likely caused the loss of a $H_2$ neutral and/or a deprotonation to account for the mass difference. However, the smaller peak at mass-to-charge 17 Da shows that a protonated (e.g. $OH^+$) ion is also generated, so it is not fully deprotonated. As such, this correlation evaporation histogram plot revealed the presence of a parent ion which was not visible in the overall mass spectrum. The post-evaporation dissociation events on the graph on the right are coming from Cu- or Zn-containing species and are not relevant to the analysis of the AA except where they can occlude peaks pertaining to molecular fragments.

Overall, the numerous Cu-related ion species due to the substrate preclude reliable identification of higher mass organic fragments, with the possible exception of the mass-to-charge peaks near 138 Da, for which the isotopic ratios do not match Cu isotopic natural abundance. Specifically, there is one much more prominent mass peak that does not have a secondary peak with would be consistent with Cu-containing ion species. Since there is some evidence from the correlated evaporation histogram for lysine on Cu of a possible organic ion at 69 Da in the 1+ charge state, an ion at mass-to-charge 138 Da could be related. The correlated evaporation histogram for **SI 9**, lysine on Cu which was pre-cleaned with nitric acid, also shows an ion fragment coming from 138 Da on both axes, so there may be a related, possibly organic, fragment ion. The nitric acid may have contributed to either increased Cu bonding to lysine and/or the formation of divalent metal cations, which form complexes and could fragment in that way . Overall while there may be an organic ion present at that mass-to-charge, it cannot be reliably determined that it is in fact organic, only that it does not match Cu isotope abundance. In addition, there are numerous smaller and lighter fragments likely derived from the fragmentation of lysine, which are beyond the scope of this note in SI to analyze.

**SI 7.** Enlarged Correlation Histogram from SI 6, showing two post-evaporation dissociation tracks into mass-to-charge 69 Da on y-axis, indicating one dissociated ion at mass-to-charge 16 Da on the x-axis and another one at 17 Da.

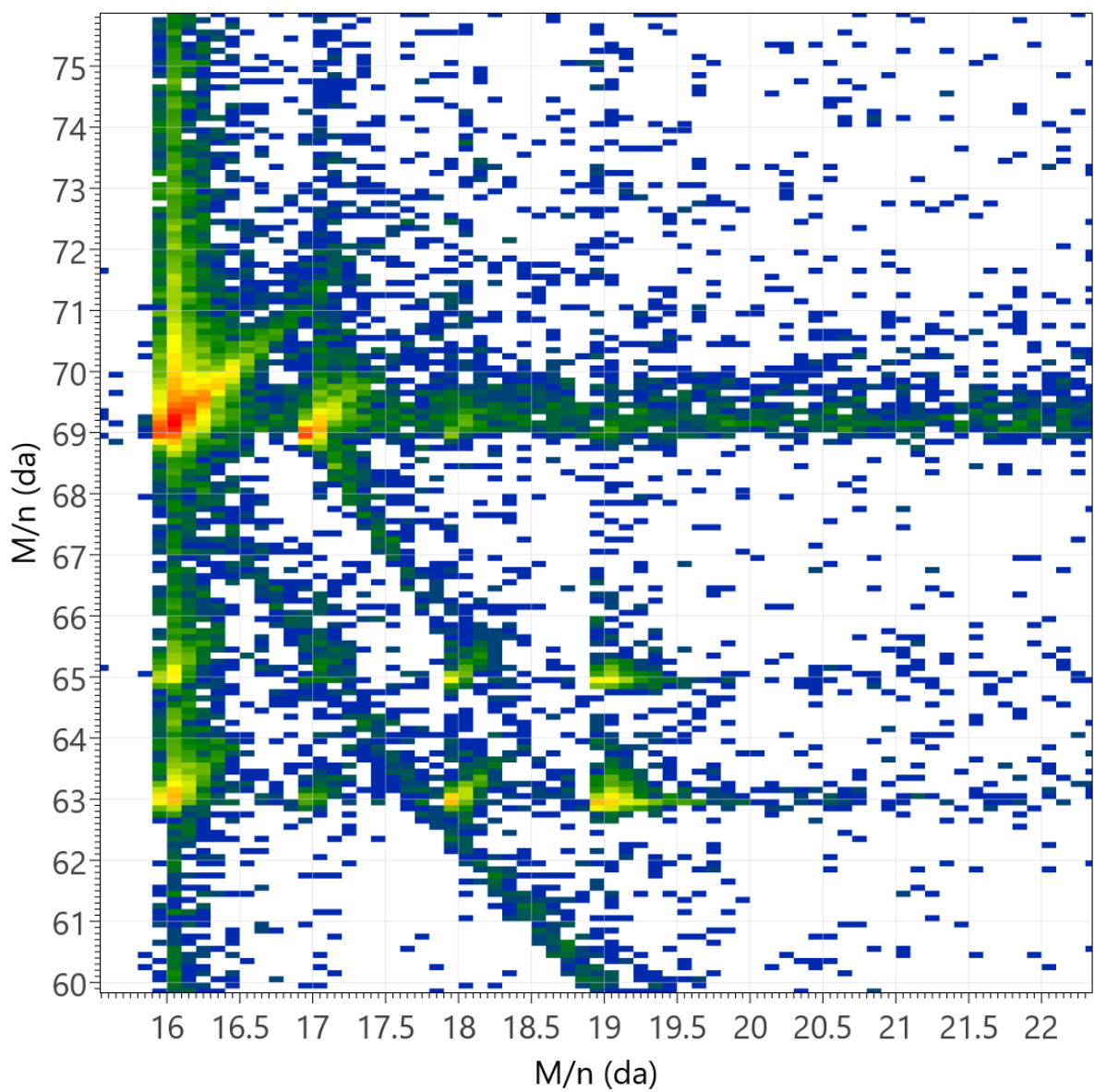

**SI 8.** Mass spectra of lysine on CuZn pre-cleaned with nitric acid, rinsed, then drop-cast. The mass spectrum demonstrates formation of CN complexes with copper and corrosion-type effects in addition to copper oxides, demonstrating highly abundant clusters.

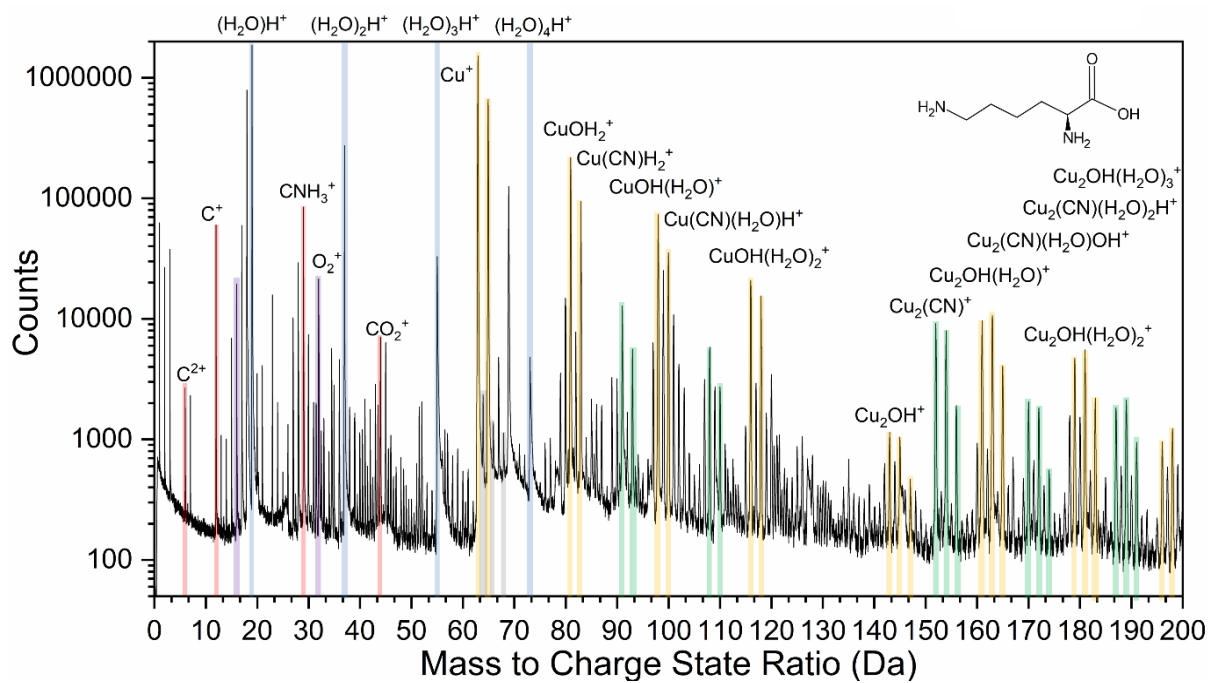

**SI 9.** Enlarged ROI cylinder map and laser powers for individual cylinders in table form

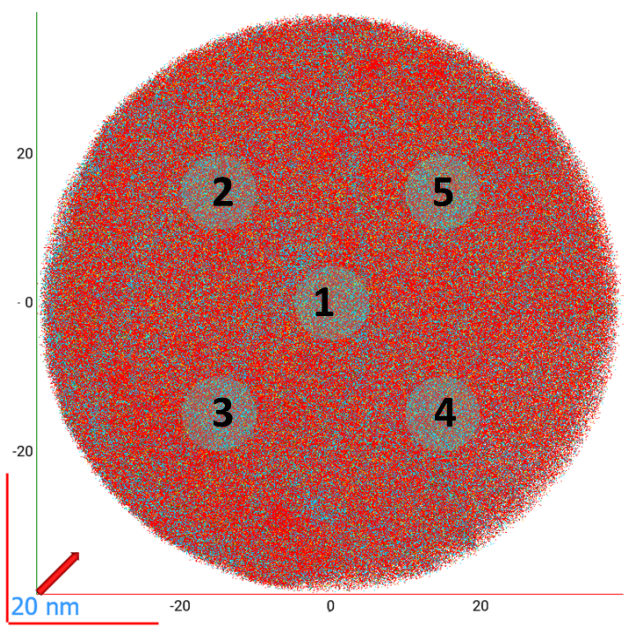

Arginine on CuZn (40M ions on two different laser pulse energies – 40pJ and 140pJ)

| 40pJ | $(H_2O)H^+$ | $(H_2O)_2H^+$ | $(H_2O)_3H^+$ | $H_2O_3/\Sigma HxOy$ ratio (CIR) |
|---|---|---|---|---|
| Cylinder 1 | 45414 | 5950 | 1632 | 32.5 |
| Cylinder 2 | 49039 | 7751 | 2318 | 25.5 |
| Cylinder 3 | 42010 | 7653 | 4785 | 8.8 |
| Cylinder 4 | 40036 | 12800 | 1430 | 37.9 |
| Cylinder 5 | 19959 | 2271 | 1391 | 16.9 |

| 140pJ | $(H_2O)H^+$ | $(H_2O)_2H^+$ | $(H_2O)_3H^+$ | |
|---|---|---|---|---|
| Cylinder 1 | 152716 | 20051 | 5600 | 31.7 |
| Cylinder 2 | 128297 | 20051 | 4237 | 36.0 |
| Cylinder 3 | 127875 | 24747 | 22420 | 5.7 |
| Cylinder 4 | 94152 | 17715 | 1150 | 98.3 |
| Cylinder 5 | 98423 | 16830 | 4017 | 29.7 |

### Note on the use of nanoporous Cu as substrate:

Interestingly, based solely on the mass spectrum, the maximum peak height of any copper oxides in that dataset relative to the $^{63}$Cu peak itself does not exceed ten percent (5.8% at in **Figure 2(c)**, although for the entire dataset 24.3% of ranged peaks were Cu (atomic %). This would suggest that while there will always be some question as to the relative contribution of those cluster ions versus organic fragments that may fall at these mass-to-charge ratios. As shown in **SI 5**, there are such clusters which occur at the most abundant masses in the arginine mass spectra in **Figure 3**. Since the peaks located approximately where the hydrated Cu oxides were located have maximum heights well in excess of the $^{63}$Cu$^+$ peak, one argument might be that such peaks were organics and not Cu oxides. However, close examination of the correlated evaporation histogram demonstrates that most of the peaks at those locations come from fragmentation. The mass-to-charge ion peaks between mass 77 - 83 Da, which are assigned as CuOH$_X$ (predominately CuOH$_2$) and between 97 – 103 Da, which are assigned as CuO$_2$H$_x$ (predominately CuO$_2$H$_3$), can be found in the lysine mass spectra without post-evaporation dissociation. However, in the arginine correlated evaporation histogram, the post-evaporation dissociation tracks show one fragment going to mass 63 or 65 (copper isotopes) or 77 – 83 Da (CuOH$_x$) respectively.

One additional problem with nanoporous Cu emerges if the surface is pre-cleaned with nitric acid and rinsed with Type 1 ultrapure DI water several times and an amino acid is then drop cast and frozen. As shown in **SI 8**, copper clusters containing up to 5 Cu atoms will form in solution, along with both copper oxides and copper cyanide (CN) ions with extra waters, up to cluster size 4. The use of nitric acid with nanoporous Cu or likely any acid pre-cleaning would therefore be contraindicated.

**SI 10.** Comparison of mass-to-charge state of r and amino acids (a) cryogenically prepared elemental S, derived from a mass spectrum of Cr-coated S (b) arginine in water on nanoporous Cu (c) cysteine, a S -containing amino acid, on Au

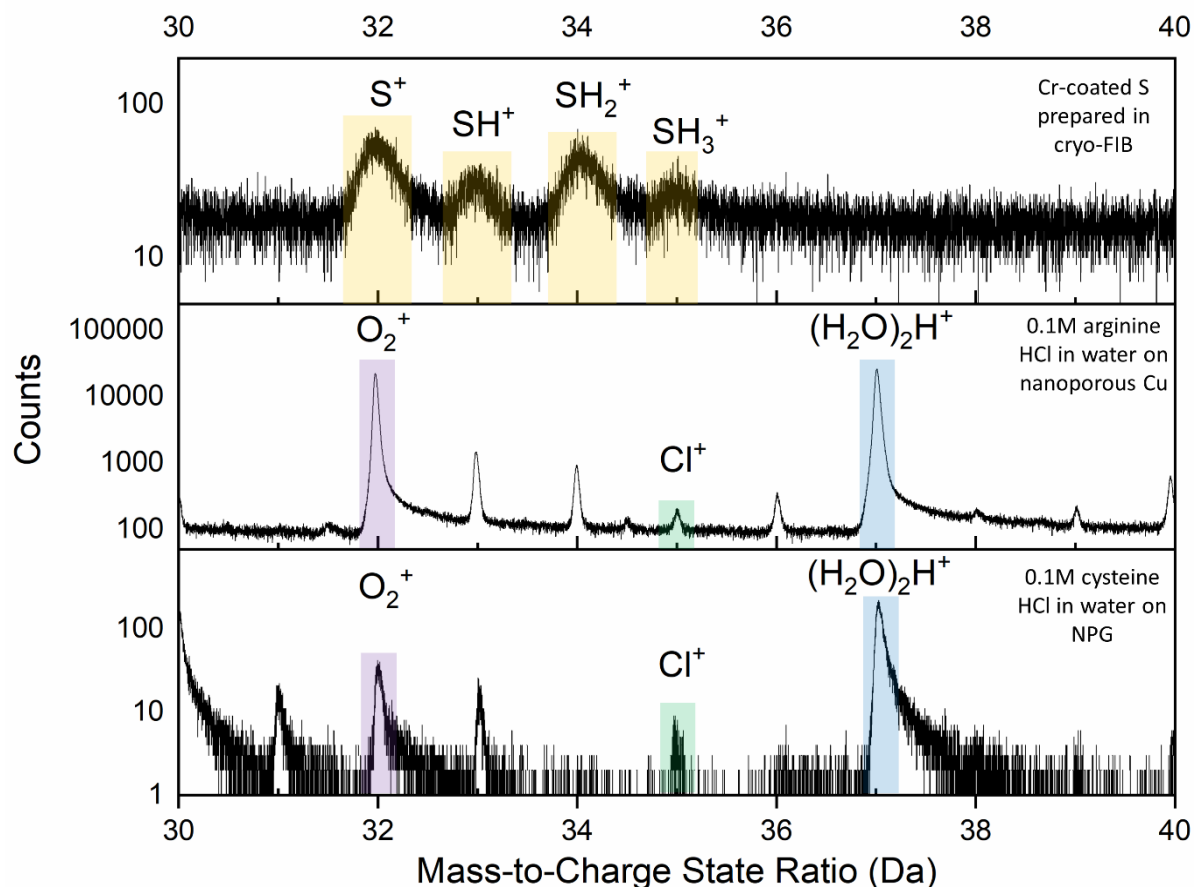